\documentclass{pasj00}

\def\deg{$\rm ^{\circ}$\ }
\def\degc{$\rm ^{\circ}C$\ }

\begin{document}
\SetRunningHead{H.Tomida et al.}{SSC onboad ISS}

\title{Solid-state Slit Camera (SSC) on Board MAXI}

%

%
\author{%
   Hiroshi \textsc{Tomida},\altaffilmark{1}
   Hiroshi \textsc{Tsunemi},\altaffilmark{2}
   Masashi \textsc{Kimura},\altaffilmark{2}
   Hiroki \textsc{Kitayama},\altaffilmark{2}
   Masaru \textsc{Matsuoka},\altaffilmark{1}
   Shiro \textsc{Ueno},\altaffilmark{1}
   Kazuyoshi \textsc{Kawasaki},\altaffilmark{1}
   Haruyoshi \textsc{Katayama},\altaffilmark{3}
   Kazuhisa \textsc{Miyaguchi},\altaffilmark{4}
   Kentaro \textsc{Maeda},\altaffilmark{4}
   Arata \textsc{Daikyuji},\altaffilmark{5}
   and
   Naoki \textsc{Isobe},\altaffilmark{6}
   }
\altaffiltext{1}{ISS Project Science Office, Institute of Space and Astronautical Science, Japan Aerospace Exploration Agency, 2-1 Sengen, Tsukuba, Ibaraki, 305-8505 Japan}
\email{tomida.hiroshi@jaxa.jp}
\altaffiltext{2}{Department of Earth and Space Science, Osaka University, 1-1 Machikaneyama, Toyonaka, Osaka 560-0043 Japan}
\altaffiltext{3}{Earth Observation Research Center, Japan Aerospace Exploration Agency, 2-1-1, Sengen, Tsukuba, Ibaraki 305-8505}
\altaffiltext{4}{Solid State Division, Hamamatsu Photonics K.K., 1126-1 Ichino, Higashi, Hamamatsu, Shizuoka 435-8558, Japan}
\altaffiltext{5}{Department of Applied Physics, Miyazaki University, 1-1 Kihanadai-nishi, Miyazaki, Miyazaki 889-2192, Japan}
\altaffiltext{6}{Department of Astronomy, Kyoto University, Oiwake-cho, Sakyo-ku, Kyoto 606-8502, Japan}


\maketitle

\begin{abstract}
Solid-state Slit Camera (SSC) is an X-ray camera onboard the MAXI mission of the International Space Station.
Two sets of SSC sensors view X-ray sky using charge-coupled devices (CCDs) in 0.5--12\,keV band.
The total area for the X-ray detection is about 200\,cm$\rm ^2$ which is the largest among the missions of X-ray astronomy.
The energy resolution at the CCD temperature of $-$70~\degc is 145\,eV in full width at the half maximum (FWHM) at 5.9\,keV,
and the field of view is 1\deg .5 (FWHM) $\times$ 90\deg for each sensor.
The SSC could make a whole-sky image with the energy resolution good enough to resolve line emissions, 
and monitor the whole-sky at the energy band of $<$ 2\,keV for the first time in these decades.
\end{abstract}

\section{Introduction}
\label{sec:intro}

The Monitor of All-sky X-ray Image (MAXI) mission (\cite{matsuoka2009}) is 
an all-sky monitor (ASM) launched by Space Shuttle Endeavor on 2009 Jul 16.
After the successful installation onto the Japanese Experiment Module - Exposed Facility (JEM-EF or Kibo EF) 
on the International Space Station (ISS) by astronauts using the remote manipulator system,
MAXI was activated on 2009 Aug 3.
The initial checkout was also successfully completed,
and MAXI is in the normal observation phase from Mar 2010.
In the 1-year observation, 
MAXI has discovered many transient phenomena in X-ray band such as X-ray bursts, 
Gamma-ray bursts, X-ray flashes, star flares, outbursts of recurrent nova.
MAXI now occupies the essential position in monitoring astronomy.

The SIS (Solid-state Imaging Spectrometer) onboard the ASCA satellite was 
the first X-ray photon counting CCD (charge-coupled device) camera in space (\cite{burke1991}).
Its energy resolution was good enough to resolve characteristic K-shell X-rays from heavy elements such as oxygen, silicon, and iron.
The ASCA/SIS opened a new window of the X-ray astronomy.
Since then, CCDs have been standard focal plane detectors of X-ray missions, 
such as Chandra (\cite{weisskopf2002}), XMM-Newton (\cite{Lumb2000}), Swift (\cite{gehrels2004}), and Suzaku (\cite{mitsuda2007}).
For the X-ray monitoring with a large field of view,
the SXC of HETE2 employed X-ray CCDs for the first time (\cite{Villasenor2003}).

MAXI has two types of X-ray camera, the Gas Slit Camera (GSC) 
and the Solid-state Slit Camera (SSC).
The SSC is a CCD camera covering 0.5$-$12\,keV energy range.
The energy band below 2\,keV has not been covered by ASMs in these decades,
which is achieved with MAXI/SSC.
The moderate energy resolution enables us to make all-sky maps with resolved line emissions.
The X-ray detection area of the SSC is unprecedentedly large to obtain the enough photon statistics.

This paper describes the design and the results of the ground performance test for the SSC.
We overview the SSC system in section\,\ref{sec:SSC_system}.
Details of the sensor unit and the onboard data processing are described 
in section\,\ref{sec:SSC_units} and \ref{sec:processing}, respectively. 
The results of the CCD screening and the calibration on the ground are summarized in section\,\ref{sec:pre_test}.
In-orbit performance of the SSC is described in Tsunemi et al. (2010).

\section{SSC system}
\label{sec:SSC_system}

\begin{figure*}
\begin{center}
\FigureFile(140mm,){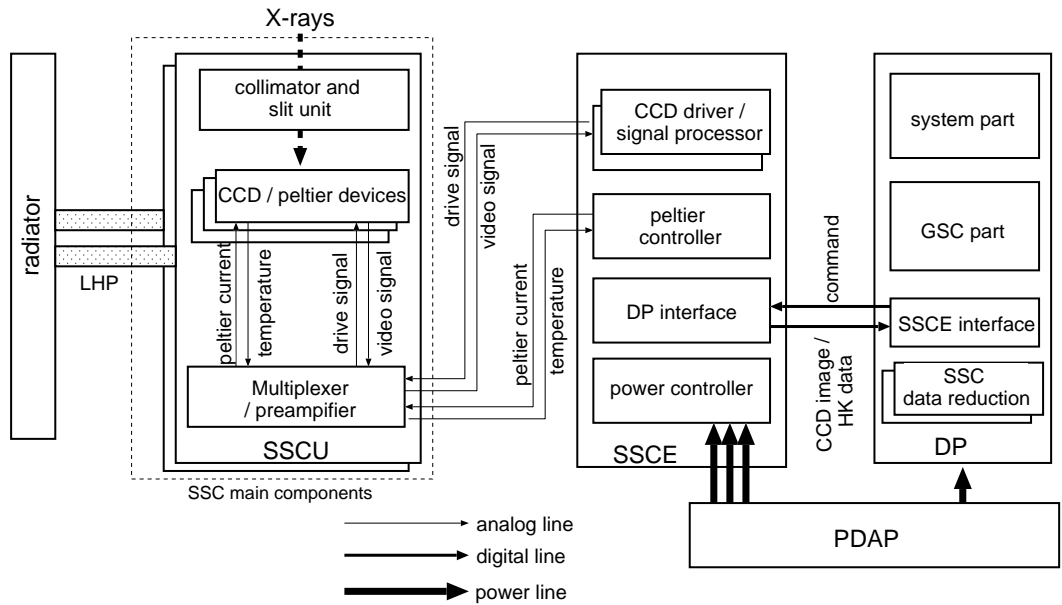}
\end{center}
\caption{Block diagram of the SSC including peripheral systems and signal flow among the components.}
\label{fig:system} 
\end{figure*}  

\begin{figure*}
\begin{center}
\FigureFile(60mm,){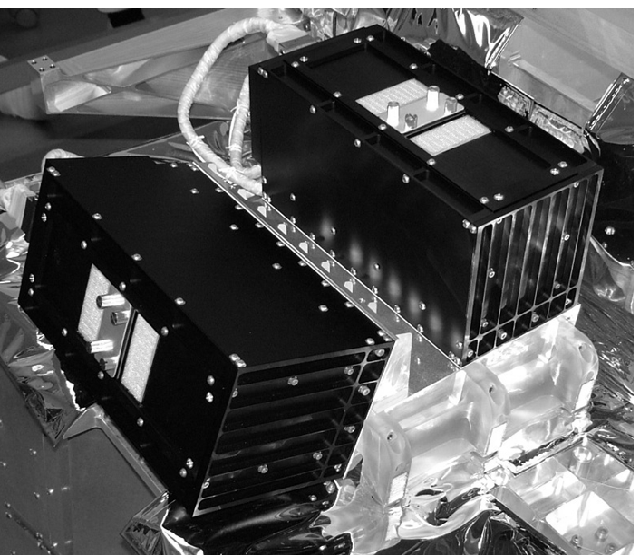}
\FigureFile(90mm,){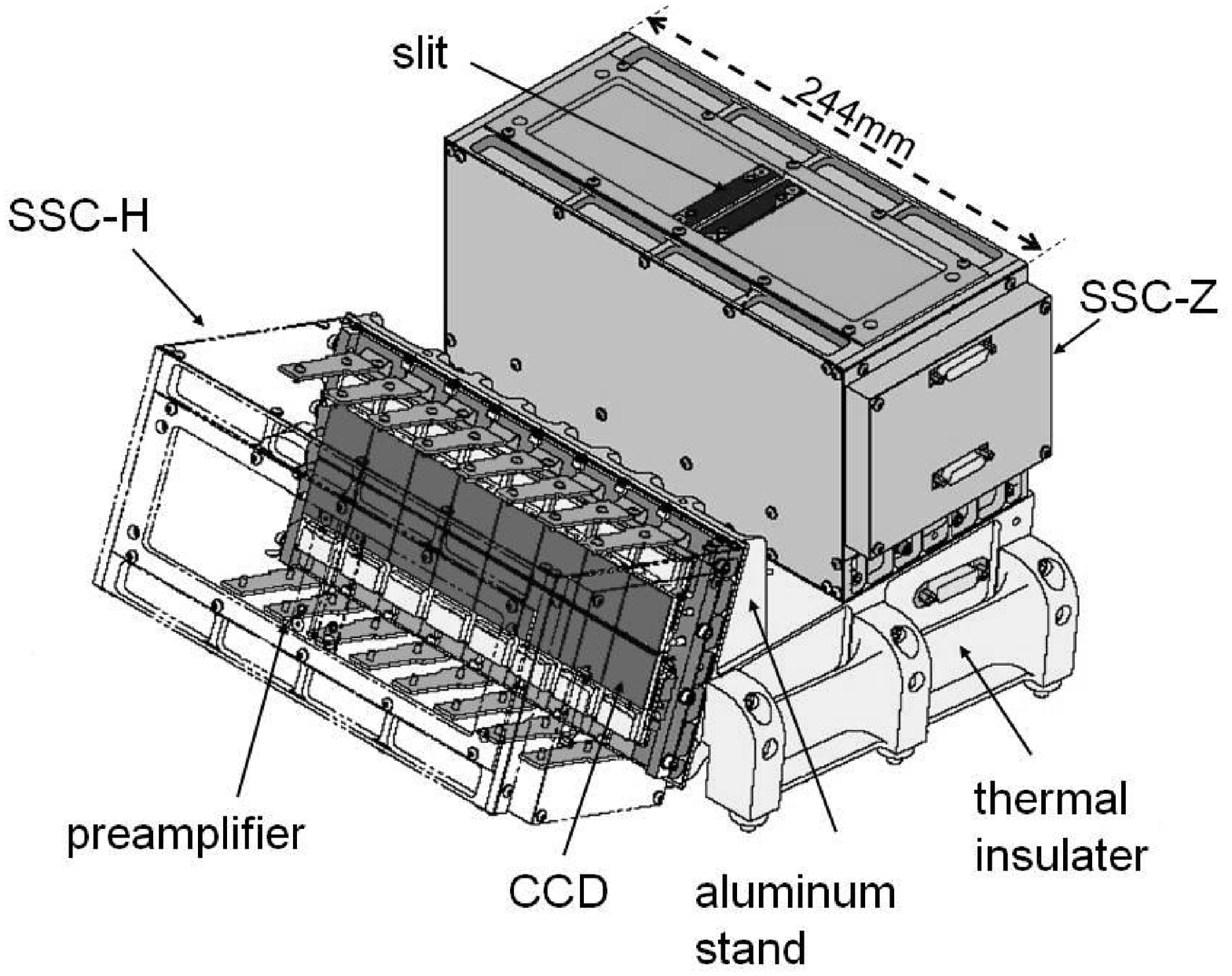}
\end{center}
\caption{ Left is a photograph of SSC units mounted on the aluminum stand before the installation to MAXI.
Slit apertures are protected by cover plates (non-flight item). 
Multi-layered insulator will cover the SSCUs in the final configuration.
Right is an exploded view of the SSC.}
\label{fig:sscu} 
\end{figure*}

Figure\,\ref{fig:system} shows the block diagram of the SSC and the peripheral systems.
The main components of the SSC are two SSC Units (SSCUs).
Figure\,\ref{fig:sscu} is a photograph and an exploded view of SSCUs.
The SSCU is a sensor part of the SSC.
One SSCU consists of CCD units, preamplifiers, multiplexers, a collimator and slit unit, and a calibration source.
The ISS orbits around the earth like the moon
in which the ISS always faces the same side towards the earth.
So MAXI is placed such that it always sees the sky direction.
Since one of two SSCUs is placed so as to monitor the zenithal sky,
it is called SSC-Z.
The other unit is called SSC-H which sees $+$20\deg above the horizontal (forward moving) direction of the ISS/MAXI.
The SSC Electronics (SSCE) controls SSCUs.
The SSCE generates CCD drive signals, and digitizes the analog signals from CCD.
The SSCE also controls CCD temperature 
using peltier devices embedded in each CCD unit.
Heat from peltier devices in SSCUs is transferred to fixed radiator panels by using a Loop Heat Pipe (LHP).
The digitized data from the SSCE are transferred to the Data Processor (DP).
The DP analyzes image data from the SSCE, extracts X-ray events, edits them into the telemetry data,
and sends them to the ground via the ISS/JEM-EF.
The DP also relays commands from the ground to the SSCE.
The Power Distributor of Attached Payload (PDAP) supplies electric power to the SSCE in three channels.
One channel is used for the CCD operation, and other two are for peltier currents in SSC-H and SSC-Z.

\section{SSC Units}
\label{sec:SSC_units}

The body of the SSCU is made of aluminum.
To suppress the light reflection,
black inorganic anodized aluminum alloys are used for the surface.
The SSCU is 244$\times$112$\times$124\,mm, and 3.9~kg including a collimator and slit unit.

\subsection{CCD Unit}
\label{sec:CCD}

\begin{figure}
\begin{center}
\FigureFile(80mm,){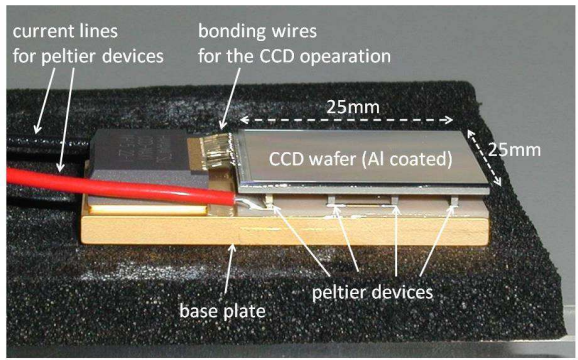}
\end{center}
\caption{ A bird's-eye view of a CCD unit in the SSC.}
\label{fig:ccd} 
\end{figure}

\begin{figure}
\begin{center}
\FigureFile(80mm,){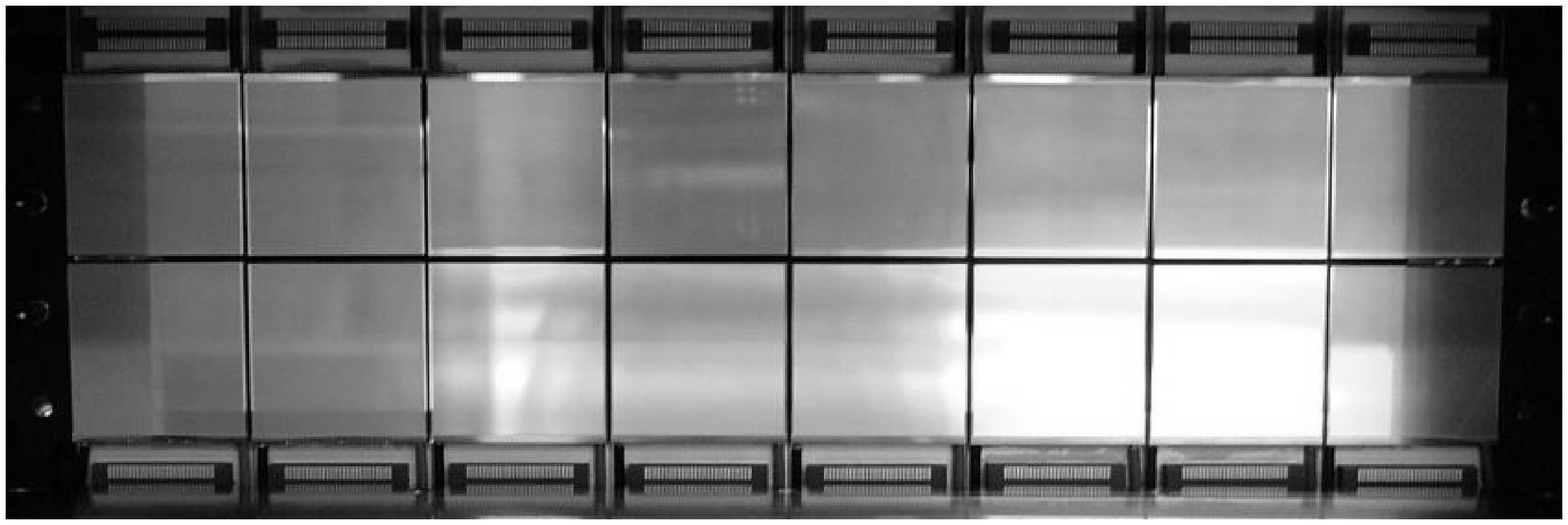} \\ 
\FigureFile(80mm,){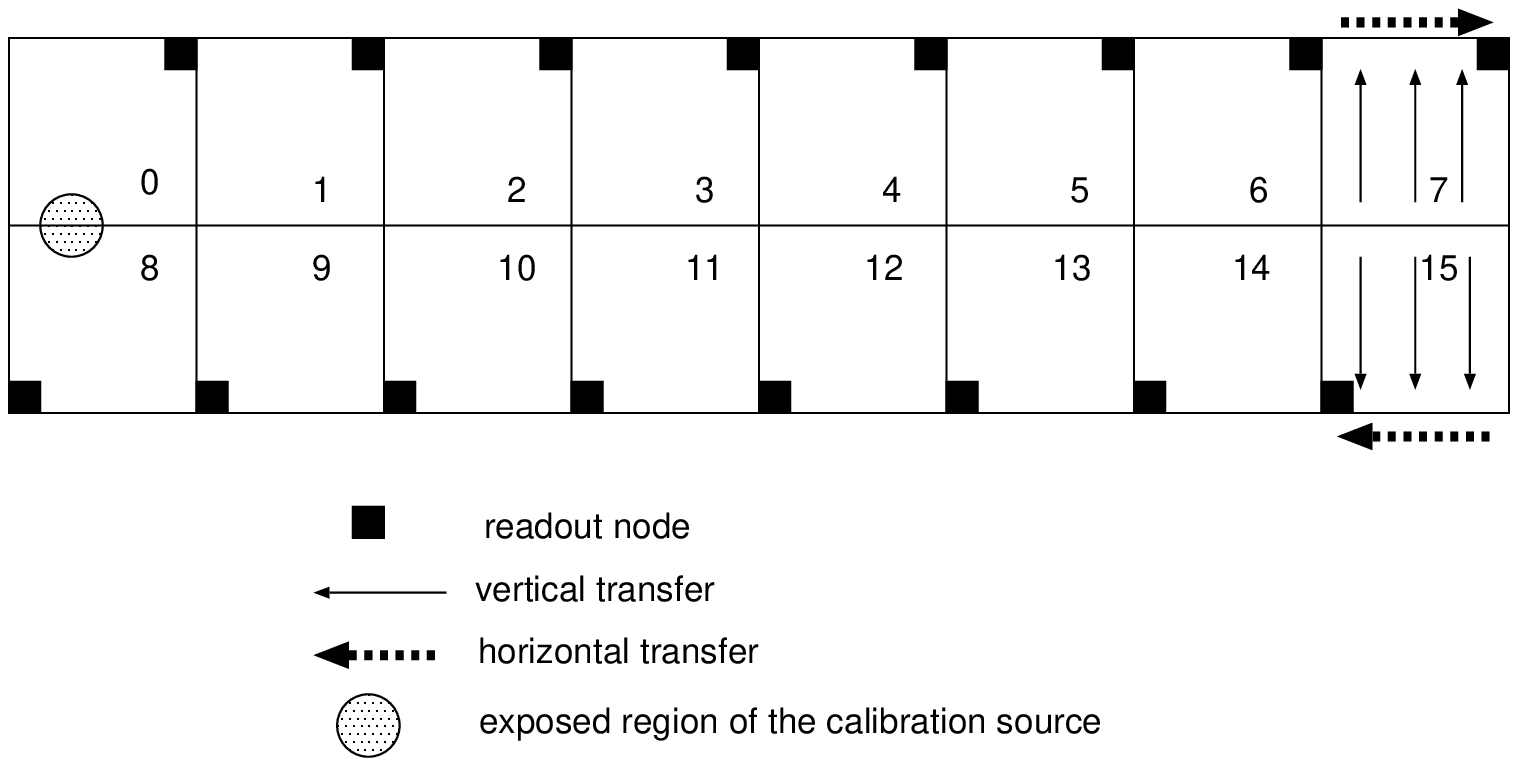}
\end{center}
\caption{ The upper panel is a photograph of CCD array in an SSCU.
The gap width between adjacent CCDs is 0.4\,mm.
The lower panel shows the position of the readout node, the direction of horizontal and vertical transfer,
and the region exposed to the calibration source. 
The numbers in the lower panel (0$\sim$15) are CCDIDs.}
\label{fig:ccd_array} 
\end{figure}

CCD units used in the MAXI/SSC are fabricated by Hamamatsu Photonics K.K.\footnote{http://www.hamamatsu.com/}
using the CCD, FFTCCD-4673.
A photograph of a CCD unit is shown in Figure\,\ref{fig:ccd}. 
The CCD unit consists of a CCD wafer, peltier devices, and a base plate made of copper-tungsten.
The CCD wafer is front-illuminated p-type CCD operated in full frame transfer mode.
The pixel number for the X-ray detection is 1024\,$\times$\,1024,
and the pixel size is 24\,$\mu$m$\,\times$\,24\,$\mu$m, 
giving the detection area of about 25\,mm\,$\times$\,25\,mm.
We employ two phase clock for the vertical and horizontal transfers.
The weight is 55\,g.
The number of pin is 32 for the CCD operation, 
and there are two other electric lines in which the peltier current of 1.2\,A flows at the maximum.
In order to obtain a large X-ray detection area,
each SSCU includes 16 CCD units that are aligned in 2$\times$8 array as shown in Figure\,\ref{fig:ccd_array}.
The total X-ray detection area of the SSC (32 CCD units) is about 200\,cm$^2$. 


CCD is sensitive for optical and infrared lights that degrade the performances of the X-ray detection. 
In order to avoid it,
aluminum of 0.2 $\mu$m thick is coated on the CCD surface,
which makes the structure of the SSCUs quite simple.
The aluminum coat is free from pinhole,
which is confirmed by illuminating the CCD with optical light.

Since the SSC does not have an X-ray mirror, 
the energy range of the SSC is determined by mainly CCD wafer.
The quantum efficiency for soft X-ray is limited by absorption at the gate structure (dead layer) 
and aluminum on the front surface of a CCD wafer.
That of hard X-ray is limited by thickness of the depletion layer.
The gate structures are made of SiO$_2$ and Si
whose designed thickness are 0.8 and 0.1\,$\rm \mu$m, respectively.
The designed value of depletion layer thickness is about  70~$\rm \mu$m.
Then the energy range determined by CCD is 0.5--15keV where the quantum efficiency is larger than 10\%. 
However, the actual energy end of the SSC coverage is limited by dynamic range of analog-to-digital converters.
Then the energy range of the SSC including the electronics is 0.5--12\,keV.

\subsection{Collimator and Slit Unit (CSU)}
\label{sec:slit_and_colliamtors}

\begin{figure*}
\begin{center}
\FigureFile(140mm,){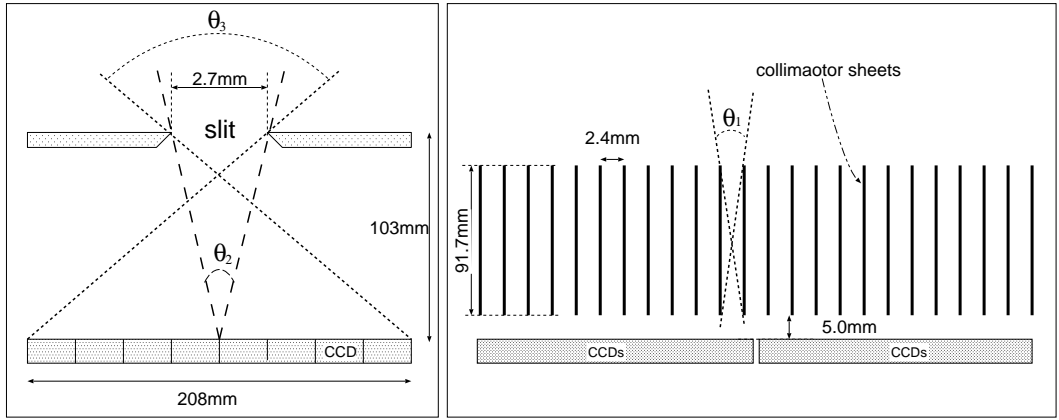}
\end{center}
\caption{Schematic view of a collimator and slit unit (CSU) in an SSCU.
The left figure shows the positional relation between the slit and CCDs.
The right figure, perpendicular to the left figure, shows the 24 collimator sheets and CCDs.}
\label{fig:col} 
\end{figure*}

Each SSCU has a collimator and slit unit (CSU).
Figure\,\ref{fig:col} is a schematic view of the CSU.
Collimators of the CSU restrict the field of view (FOV) to be a fan-beam.
The collimator sheet pitch determines the narrow FOV to be 3$^{\circ}$.0 
in bottom-to-bottom ($\rm \theta _1$ in Figure\,\ref{fig:col}).
A scan of an object with constant flux forms a triangular response,
whose full width at the half maximum (FWHM) is 1$^{\circ}$.5. 
On the other hand, the combination of 8 CCD units and the slit 
determine the long FOV to be 90$^{\circ}$ ($\rm \theta _3$ in Figure\,\ref{fig:col}).
One orbit scan corresponds to 90$^{\circ}$ $\times$360$^{\circ}$.
About 29\% of the entire sky are not covered by SSC in an orbit scan.
It takes about 70 days to cover the unobservable region, 
depending on the precession period of the ISS orbit plane.
In addition, SSC cannot observe the region around the Sun,
which is too bright to block optical light with the aluminum coat on the CCD surface.
Then it takes about a half year to obtain the actual all-sky image.

The collimator consists of thin sheets made of phosphor bronze with the thickness of 0.1~mm. 
There are 24 sheets with 2.4~mm interval that are placed 5.0\,mm above the CCD units. 
The surface of the collimator sheet is chemically etched to suppress the  X-ray reflection,
and plated with black chromium to suppress the optical light.

The slit of the SSC consists of two tungsten bars with sharp edges. 
The width between the two edges is 2.7\,mm.
The angular resolution depends on the acquisition angle
which is the angle between the incident X-ray direction and the normal of the slit plane.
For larger acquisition angle, angular resolution becomes better,
while the effective area becomes smaller by a cosine factor.
At the acquisition angle=0$^{\circ}$.0, 
the angular resolution ($\rm \theta _2$ in Figure\,\ref{fig:col}) is 1$^{\circ}$.5.

\subsection{Calibration Source}
\label{sec:cal_source}

Each SSCU has an $^{\rm 55}$Fe calibration source.
The calibration source is mounted on the CSU
so that small part of 2 CCDs (CCDID=0,8) in each SSCU is exposed to
Mn K$\alpha$(5.9~keV) and K$\beta$(6.5~keV) X-rays.
The circle area in Figure\,\ref{fig:ccd_array} shows the exposed region.
The main purpose of the calibration source is to monitor the long-term trend of the gain.
The fluxes are high enough to determine the gain with the statistical error of 0.1\,\% in one-day accumulation.

\subsection{Thermal Design}
\label{sec:sscu_thermal}

The energy resolution and the detection limit at the lower energy region of the SSC are limited by various conditions,
one of which is a variance of thermally excited charges (electrons) in CCD pixel during the exposure and the charge transfer. 
The low temperature operation is essential to suppresses the thermal noise.
As for the SSC, one-stage peltier devices are used to cool CCDs.
Each CCD unit is equipped with peltier devices as shown in Figure\,\ref{fig:ccd}.
The CCD wafer in each unit is mechanically supported by 12 pairs of peltier devices (24 posts made of bismuth telluride).
There is no other mechanical support, which suppress the heat input through the conduction.
Since eight CCD units are serially connected in the peltier line, 
there are four electric current lines in SSC,
each of which can be controlled independently.
For the current control,
we have two operation modes; the current constant mode and the temperature constant mode.
In the former mode, the constant electric current is provided into the peltier line.
In the latter mode, the peltier current is controlled so that the CCD temperature is kept at the target temperature. 

The electrical resistance of the peltier devices in each CCD unit is about 1\,$\Omega$,
and the peltier current in standard SSC operation is about 1\,A.
Hence, peltier devices of 32 CCDs generate 32\,W.
In addition, 5~W is required to operate analog circuits in SSCUs. 
Then the power consumption of the SSCUs is 37~W
that is transferred through the LHP
to radiator panels on MAXI.
It is designed that the LHP and radiator system could cool SSCUs to around $-$20\degc .
The hot sides of peltier devices are thermally connected to the SSCU body.
The peltier devices give the temperature difference of $>$40\,\degc between the CCD wafer and the SSCU body.
Hence, CCDs are to be operated below $-$60\degc.
The body of SSC-H and SSC-Z are, both, placed on an aluminum stand, then they are thermally combined to each other.
The aluminum stand is mounted on the GSC-H unit with a thermal insulator made of polycarbonate (Figure\,\ref{fig:sscu}).

The total consuming power of SSC is 90~W in typical;
26~W for CCD operation and 64~W for the CCD cooling. 
21~W of 26~W for the CCD operation is consumed in the SSCE,
and 5~W is in SSCUs.
32~W of 64~W for the CCD cooling is used in the SSCE,
and the rest 32~W is for peltier devices in the SSCUs.
Then SSCE including DC/DC converters generates 53~W.
The SSCE is maintained at about 20~\degc by the fluid loop system on the ISS/JEM.

%
%

\section{Onboard Data Processing}
\label{sec:processing}

\subsection{CCD drive}
\label{sec:CCDdrive}

CCD clocking signals are generated in the SSCE.
CCD has two-dimensional pixel array,
while the SSC requires only one-dimensional position information since the SSC conducts a scan observation.
Hence, we operate the SSC in parallel-sum mode,
that is, charges in multiple pixels are summed in serial register at the bottom of the imaging region.
The summed charges in the serial register are horizontally transferred to the readout node.
The number of summed pixels in observation can be selected from 
16, 32, 64 by commands.
We call this number as the ``binning'' parameter 
that is selected to be 64 for the standard observation.
The larger binning gives the better time resolution,
that provides better angular resolution in X-ray sky map.
The binning of 1, 2, 4, 8 is used only for diagnosis. 
The horizontal pixel clock is 8\,$\rm \mu$s pixel$\rm ^{-1}$,
and the vertical clock is 100\,$\mu$s line$\rm ^{-1}$ for binning=1.
The video signal from 16 CCDs in a SSCU is processed serially
by one readout electronics.
When one CCD is readout, other 15 CCDs receive no clocking signal and are in the exposure state.
Since it takes 0.232\,s to readout one CCD in binning=64, 
the read cycle for 16 CCDs is 3.719\,s.
Clocking voltages are common for 16 CCDs in each SSCU.
During the exposure state, CCDs are left in the flip mode (Miyata \etal\ 2004),
in which the bias voltage of CCDs is periodically inverted from +3\,V to $-$9\,V to reduce the dark current.

\bigskip


In the ISS orbit, CCDs are exposed to high flux of charged particles,
which degrades the CCD performance.
We expect the increase of the CTI, dark current, and the number of hot pixels.
The degradation of the MAXI/SSC was estimated based on the experiments (\cite{miyata2002} and \cite{miyata2003}).
It is known that the charge injection (CI) technique was a good way to restore the CTI (\cite{tomida1997}).
Uchiyama et al (2009) demonstrated that the increased CTI due to charged particles 
could be restored in the orbit operation of the Suzaku/XIS with CI, and constructed a new method of correcting pulse heights.
We, then, added the CI function to the SSC, and could select by command whether or not the CI function is applied.

The CI function has two important parameters: the amount of charge and the injection period 
that is determined from the binning parameter.
In the case of binning=N (N$>$1), charge is injected at every N rows.
The signal charges in N-1 rows that include no injected charge are summed and readout in serial register.
Whereas the injected charges are transferred and discarded at the readout node.
Since the extra horizontal transfer is required in CI, the readout cycle of 16 CCDs is 5.865\,s, 
which is 1.577 times longer than that without CI.

The charge amount can be roughly controlled by commands through the clocking voltage.
The voltage can be changed in 0.1~V step, 
which gives the difference of about 6000 electrons pixel$^{-1}$ for the injected charge.
Since the larger amount of charge gives the better restoration of the CTI (\cite{tomida1997}),
we set the injected charge as much as possible with the condition that
there is no overflow during the transfer.
The charge amount is estimated to be about 10$^{5}$ electrons pixel$^{-1}$.
The CTI before the launch is negligibly small, 
so we could not confirm the CI effect in the ground test. 
The Suzaku/XIS has a capability to control the amount of injected charge precisely,
which is utilized to measure the CTI  (\cite{nakajima2008}, \cite{ozawa2009}).
The SSC, however, cannot precisely control the charge amount.
We use the CI method only for the CTI improvement.

\subsection{Digital Processing}
\label{sec:DP}

The CCD images digitized in the SSCE are sent to the DP.
We adopted the data reduction procedure used in the timing mode of Suzaku/XIS (\cite{koyama2007}).
The dark level, which is the time-averaged pixel data with neither incident X-rays nor charged particles, 
is determined for individual pixels and up-dated at every readout time.

The DP searches for the charge deposit pattern of signals,
which is called an ``event''.
The pixel data after the dark-level subtraction are called ``pulse height (PH)''.
The event search is done by referring to the PH.
An event is recognized when both of the following conditions are satisfied :
(1) a pixel has a PH between the lower and the upper thresholds defined by command, 
(2) the PH is a local maximum among adjacent three pixels in the row.

The MAXI telemetry data are transferred from the DP to the ground 
through two physical networks in the ISS/JEM-EF.
One network is MIL-STD-1553B and the other is Ethernet.
MIL-STD-1553B has higher reliability and more real-time connection between ISS/JEM-EF and the ground station than Ethernet.
However, the data transfer rate of MIL-STD-1553B allocated for MAXI is 50\,kbps (\cite{ishikawa2009}),
while that of the Ethernet is 600\,kbps.
Therefore, we designed that Ethernet data of the SSC are used for the detailed analysis (spectroscopy and diagnosis of sensors)
while the MIL-STD-1553B data are to search for the transient phenomena quickly.

\begin{figure}
\begin{center}
\FigureFile(80mm,){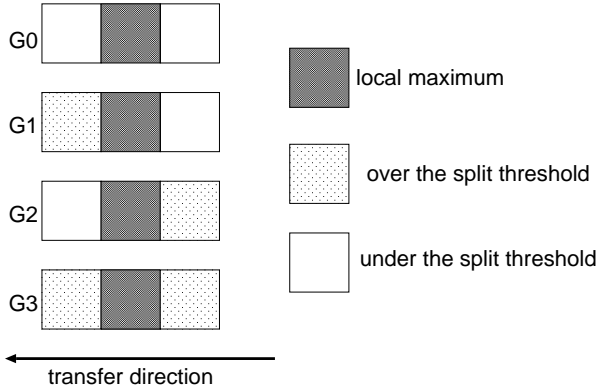}
\end{center}
\caption{Four grade patterns (G0 through G3) used in the SSC data processing.}
\label{fig:grade} 
\end{figure}

Each event data of Ethernet telemetry includes X- and Y-address (X-ray detected pixel position in CCD), 
and the PHs of adjacent 5 pixels (local maximum, two trailing pixels, and two leading pixels).
X-address is used to determine the position of X-ray in the FOV,
Y-address is used to correct PH with CTI.
If the number of event is too large,
the event data are compressed automatically or by command.
In the compress mode,
PHs of five pixels are replaced with a summed PH according to the pixel pattern called grade.
There are four grades (G0, G1, G2, and G3), which is the same to the timing mode of the Suzaku/XIS.
Figure\,\ref{fig:grade} shows the four grade patterns.
G0 is a pixel pattern that all signal charge (electrons) generated by an incident X-ray photon is confined in one pixel.
PH of the pixels adjacent to the local maximum is smaller than the value called split threshold. 
G1 and G2 indicate that either the leading pixel or the trailing pixel to the local maximum has more charge than the split threshold.
G3 is that both of the leading pixel and the trailing pixel to the local maximum have more than the split threshold.
The replaced PH of the telemetry in the compress mode is the sum of local maximum and those over the split threshold. 
In the compress mode, we can select events to be downlinked according to the grade pattern.
Events of G3 are discarded in normal operation since G3 events are thought to be generated by charged particles.

For the MIL-STD-1553B telemetry, the DP compresses every X-ray event into 16 bits.
14 bits are assigned for X-address and the summed PH, and 2 bits for the grade.
The grade definition is the same to that of Ethernet,
but the grade selection to be sent to the ground can be selected independently from the Ethernet data.
The bit assignment for X-address and PH is determined by command.
In the standard operation, 6 bits are assigned for X-address and 8 bits for PH,
where the accuracy of the position determination (X-address) is 0.36\,mm and that for the energy is 60~eV.
The number of SSC events for MIL-STD-1553B telemetry is limited to be less than 255\,events for every 16\,lines.

\section{Screening Test and Pre-flight Calibration}
\label{sec:pre_test}

\subsection{CCD screening}

Before assembly of the SSCUs,
we had many candidates of CCD units to be installed.
Then we conducted the screening test to select the CCD units suitable for the SSC.
CCDs were driven by using the $E$-$NA$ system (\cite{miyata2001}), 
and cooled down with a pulse tube cooler in a vacuum chamber.
The peltier devices were not activated.
The CCD units for the flight camera are selected from 64 candidates provided by Hamamatsu Photonics K.K.
Miyata et al (2004) described the detailed setup of the experiment and the initial results of 25~CCDs. 
We report the final results here.

At first, we determined the proper voltages to drive the CCDs.
There are 16 voltages for the operation of the MAXI-CCDs.
The 16 CCDs in each SSCU have to be operated with common voltages
since we have only one set of electronics for each SSCU.
To search for the proper voltages, the CTI and the CI ability are mainly evaluated.
From the experiment, we confirmed that 14 of 16 voltages could be common for 2 SSCUs to obtain the low CTI and good CI performance, 
while we could not operate with common voltages for two which are used to inject charge correctly.
They are ISV (Input Source for the Vertical transfer) and IGV2 (second Input Gate for Vertical transfer),
which are important to control CI.
The structure of the charge injection gate is described in the technical note of the manufacturer 
\footnote{http://sale.hamamatsu.com/assets/applications/\\
SSD/fft\_ccd\_kmpd9002e06.pdf}.
In the case of ISV, we found that CCDs were separated into two groups;
one is that appropriate voltage of ISV is 3.0\,V while the other is 3.8\,V.
We, then, decided that CCDs in the 3.8\,V group were installed to SSC-H,
and the 3.0\,V group were installed to SSC-Z.
The appropriate voltages of IG2V are scattered and cannot be properly grouped.
So we designed the SSCE such that the IG2V voltage could be set separately for each CCD.

After the operating voltages were fixed,
we measured various parameters as a function of temperature; the dark current, the number of the hot pixel, 
the number of the dead column, the readout noise, and the energy resolution.
The CTI was also examined again. 
Radio active $\rm^{55}$Fe sources were employed for the experiments.
The binning of 1 and 64 are tested.
We sorted the flight devices based on the energy resolution,
then we selected 16 CCD units having the best energy resolution for SSC-H and SSC-Z, respectively.
The readout noise of the selected CCDs ranges from 6 to 10\,electrons in root mean square (RMS) that includes system noise. 
The readout noise is measured from the fluctuation of PH in horizontal over-clocked region.
The values of CTI are $<$ 1.2$\times$10$\rm ^{-5}$ for all the selected CCDs,
where the CTI is defined as the ratio of the lost charge to the original one in a single vertical transfer. 

The hot pixels and the dead columns are troublesome 
since the fluctuation of the dark current is so large depending on the temperature. 
We, then, check the number of the hot pixels and the dead columns for the selected CCDs.
We defined the hot pixel that the PH with no X-ray is significantly higher than that of the surrounding pixels.
The dead column is defined that the event number is smaller than that of neighbor columns by 5\,$\sigma$ level
when X-ray from $^{55}$Fe are uniformly illuminated.
The numbers are evaluated in binning=1.
The total number of dead pixels and dead columns in 32 CCDs were 37 and 15, respectively. 
About half of the CCDs are free from dead pixels and dead columns.
We found that the cosmetics of the MAXI-CCDs is excellent.
 
\begin{figure}
\begin{center}
\FigureFile(80mm,){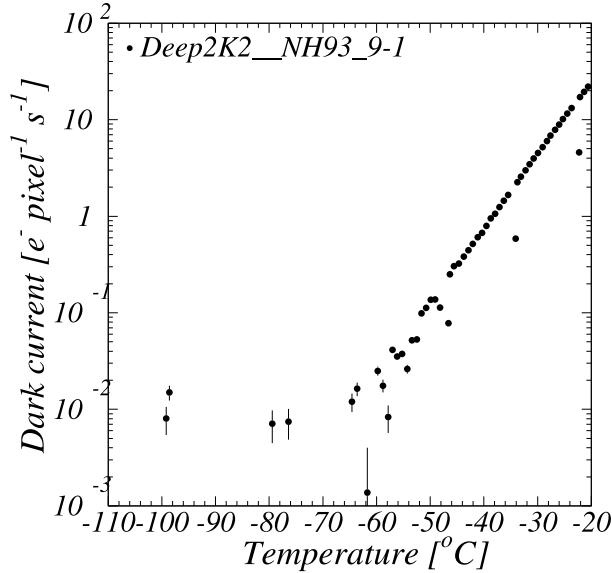}
\end{center}
\caption{The dark current as a function of the CCD temperature at the screening experiment.
The CCD unit employed is assigned to CCDID=3 in SSC-H.}
\label{fig:dark_vs_temp} 
\end{figure}

Figure\,\ref{fig:dark_vs_temp} exhibits the dark current as a function of the CCD temperature.
From the thermal analysis, the CCD temperature in the normal operation in the ISS is 
estimated to be about $-$60\,\degc (Section\,\ref{sec:sscu_thermal}).
The dark current at $-$60\,\degc is about 0.02 electron sec$\rm^{-1}$ pixel$\rm^{-1}$ in binning=1.
In the normal operation, the binning parameter is 64, 
and the readout cycle time is 5.9\,s with CI-on.
The dark current is, then, about 8 electron pixel$\rm^{-1}$ cycle-time$\rm^{-1}$,
which gives us the noise of about 3 electrons as a fluctuation in typical. 
This is smaller than the readout noise,
then we concluded that $-$60\degc was low enough to operate the CCDs at the beginning of the in-orbit operation.

Since the X-ray intensity of $^{55}$Fe sources utilized in the above performance tests was calibrated well,
we could estimate the detection efficiency at 5.9~keV.
Our experiments showed that the detection efficiency of the selected CCD units is $>$89\% at least,
while the designed depletion layer thickness of 70$\mu$m provides the detection efficiency of about 91\%.
We, hence, confirmed the detection efficiency is high enough to cover the energy range up to 12~keV.
In order to conduct the detailed spectrum analysis,
the calibration at the low energy range is also important.
We plan to measure the detection efficiency from 0.5 to 12~keV using the Crab nebula in the in-orbit operation. 

\subsection{Energy Calibration}
\label{sec:energy_cal}

\begin{figure}
\begin{center}
\FigureFile(80mm,){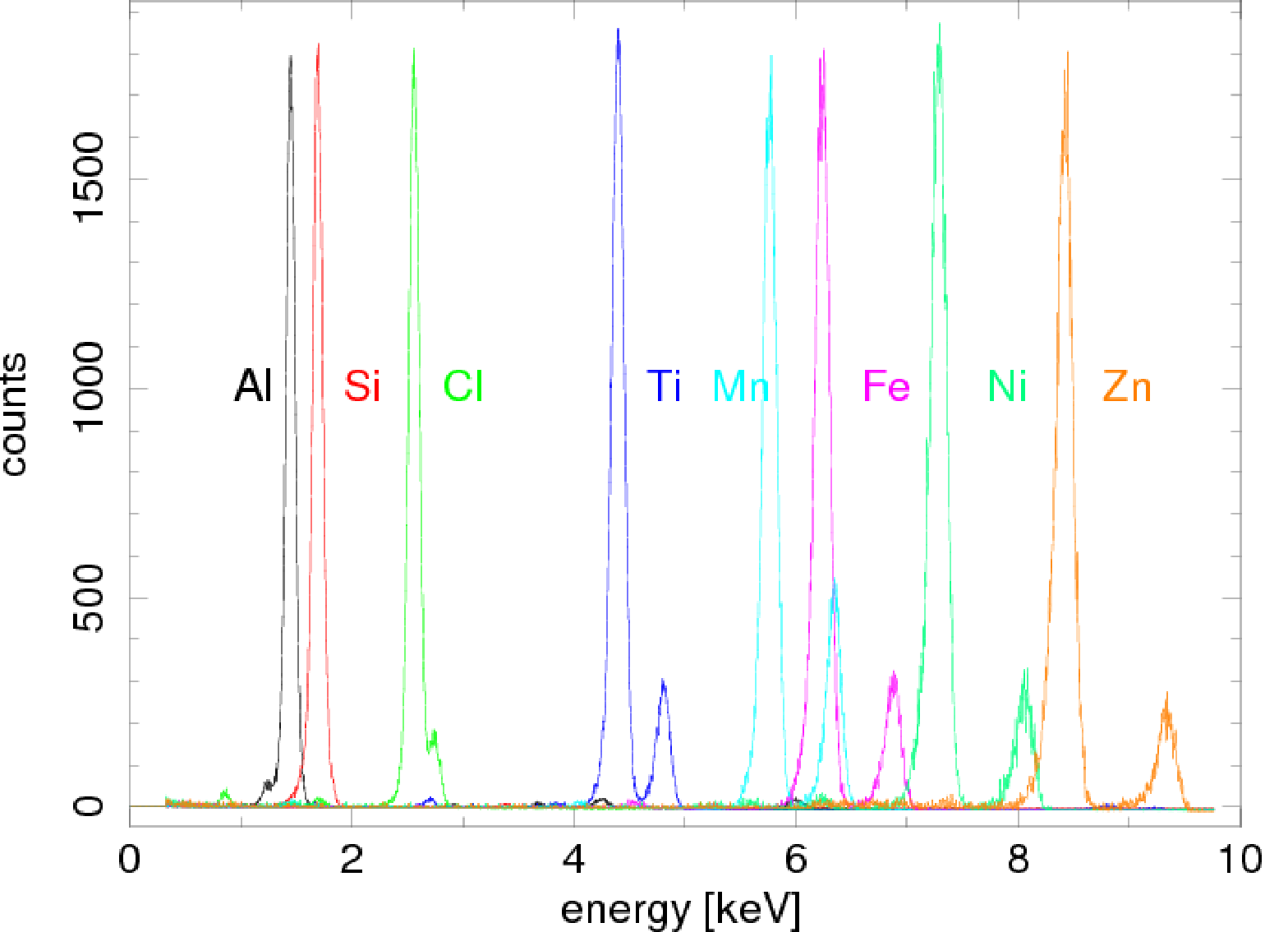}
\FigureFile(80mm,){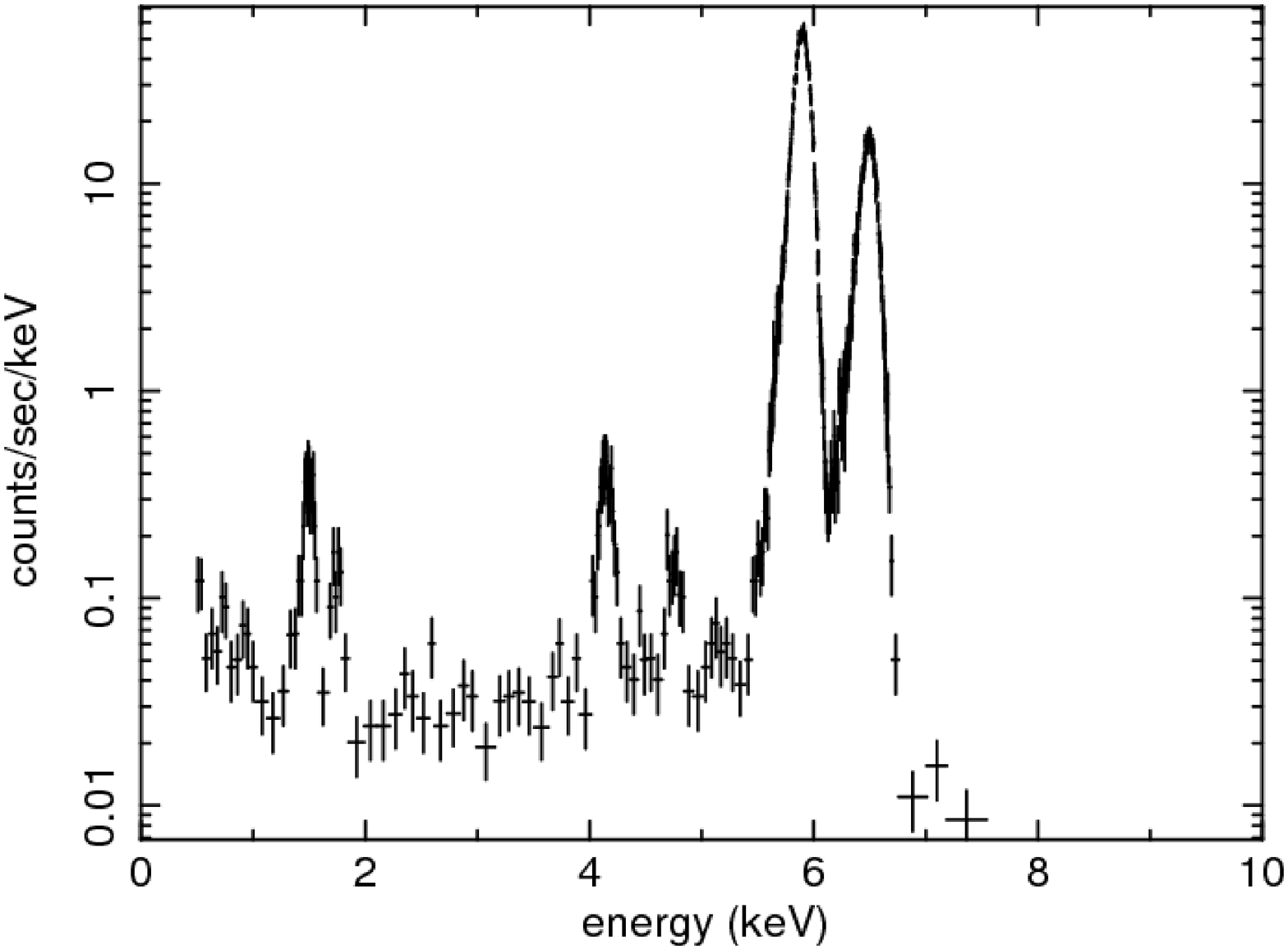}
\end{center}
\caption{
Pulse heights distribution of the SSC-H/CCDID=3.
In the top figure, fluorescent X-rays from Al, Si, Cl, Ti, Mn, Fe, Ni, Zn are irradiated.
The vertical axis is in linear scale.
In the bottom figure,
X-rays from $^{55}$Fe (5.895 and 6.490 keV) are irradiated.
The vertical axis is in log scale, and the unit is count sec$^{-1}$ keV$^{-1}$.}
\label{fig:spectrum} 
\end{figure}

The energy calibration of CCD performance with the SSCE was done before the SSC was installed into the MAXI flight structure.
Fluorescent X-rays from nine materials and Mn-K X-rays from $^{55}$Fe sources were used 
to study the energy response of the SSC.
The energy range is 0.52 (oxygen) to 12.5\,keV (selenium).
Figure~\ref{fig:spectrum} (left) shows a energy spectrum of various fluorescent X-rays.
The configuration of the calibration experiment was: 
(1) The SSCU to be calibrated is in the vacuum chamber and the other SSCU and the SSCE are in the atmosphere (room temperature at 1.0\,atom).
(2) The CSU are removed.
(3) CCDs are driven with the same operation condition as that in the orbit. 
The binning is 64 and the CI is on.
(4) The SSCU in the vacuum chamber is cooled to around $-$30\degc and peltier devices are controlled so that CCDs are at the constant temperature.
The pulse height distribution, energy resolution, 
and the linearity of the energy scale are investigated in detail.

\subsubsection{Pulse Height Distribution}
\label{sec:spectrum}

Figure\,\ref{fig:spectrum} (right) shows the spectrum of $^{55}$Fe sources with SSC-H/CCDID-3 at $-$70\degc.
The binning parameter is 64.
The spectrum is created from G0 events only.
We can see that Mn K${\alpha}$ and K${\beta}$ lines are clearly resolved,
but the spectrum shape cannot be represented by two Gaussians for Mn-K$\alpha$ and K$\beta$.
Then we fit the spectrum with a model employed in the Suzaku/XIS (\cite{koyama2007})
that includes six components : a main Gaussian, a sub Gaussian, a triangle component, a silicon escape peak,
a silicon peak, and a constant component.
We, however,  could not determine the parameters of the triangle component since this component is very small.
In this way, we found that the SSC data could be represented well with other five components.
The increase below 1.0\,keV  in Figure\,\ref{fig:spectrum} is not created by X-ray but by a background component.
The count rate of the component is about 0.1\,counts sec$^{-1}$ keV$^{-1}$ at 0.5\,keV, 
while the particle background in the orbit is 0.3\,counts sec$^{-1}$ keV$^{-1}$ 
even at the high cut-off-rigidity region (\cite{tsunemi2010}).
Although the origin of this component is still unclear,
we concluded that there is no need to pay attention to the low energy component
by applying the normal background subtraction method.

\subsubsection{Energy Resolution}
\label{sec:resolution}

\begin{figure}
\begin{center}
\FigureFile(80mm,){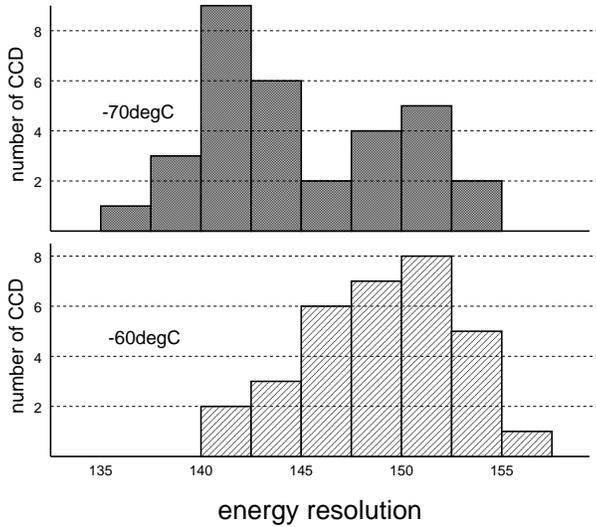}
\end{center}
\caption{Distribution of the energy resolution (FWHM at 5.9~keV) for 32 CCDs in SSCUs at $-$70\,\degc (upper) and $-$60\,\degc (lower).}
\label{fig:resolution_dist} 
\end{figure}

We defined the energy resolution as the width of the main Gaussian peak.
The energy resolution of Figure\,\ref{fig:spectrum} is 136 eV at 5.9 keV in FWHM.
Figure\,\ref{fig:resolution_dist} shows the distribution of the energy resolution
of 32 CCDs at $-$60\,\degc and $-$70\,\degc.
The average of the energy resolution is
149\,eV for $-$60\,\degc and 145\,eV for $-$70\,\degc.
Since we apply the same clocking pulses to all the CCDs in each SSCU,
we tuned them so that the averaged performance (energy resolution) becomes the best.

\begin{figure}
\begin{center}
\FigureFile(80mm,){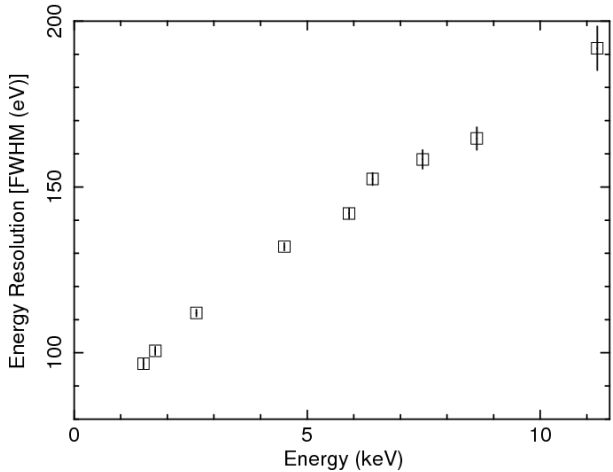}
\end{center}
\caption{Energy resolution of SSC-H/CCDID=3 as a function of X-ray energy.}
\label{fig:eres_vs_energy} 
\end{figure}

The energy resolution of the CCDs depends on the energy of incident X-rays.
Figure\,\ref{fig:eres_vs_energy} shows the energy resolution of SSC-H/CCDID-3 at $-$70\degc as a function of X-ray energy. 
G0 events are used for the spectrum analysis. 
The readout noise of 32 CCDs ranges from 5 to 10 electrons in RMS,
which is the same level to that measured in the CCD screening.

\subsubsection{Energy Scale Linearity}
\label{sec:linearity}

\begin{figure}
\begin{center}
\FigureFile(80mm,){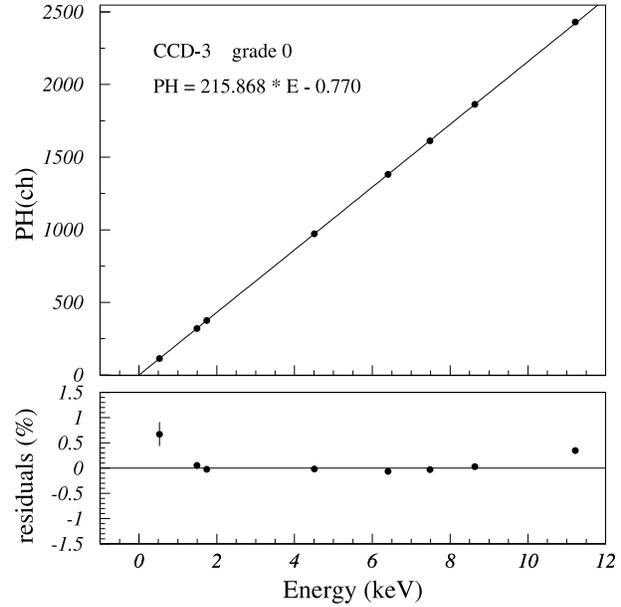}
\end{center}
\caption{Pulse height of SSC-H/CCDID=3  as a function of X-ray energy.
The lower panel shows the residuals between the best fit model 
(linear function) and the data.}
\label{fig:gain} 
\end{figure}

Figure\,\ref{fig:gain} shows the PH peak of SSC-H/CCDID-3
as a function of incident X-ray energies,
where PH peak is the center of the main Gaussian of G0 events.
We fitted the data with a linear function,
and the best fit line is shown in Figure\,\ref{fig:gain}.
We can see that the relation is well reproduced by a linear function.
No significant deviation can be seen around the silicon K-edge.
The residuals is less than 1.0~\% .

\subsection{Performance of Peltier Devices}
\label{sec:peltier_performance}

\begin{figure}
\begin{center}
\FigureFile(80mm,){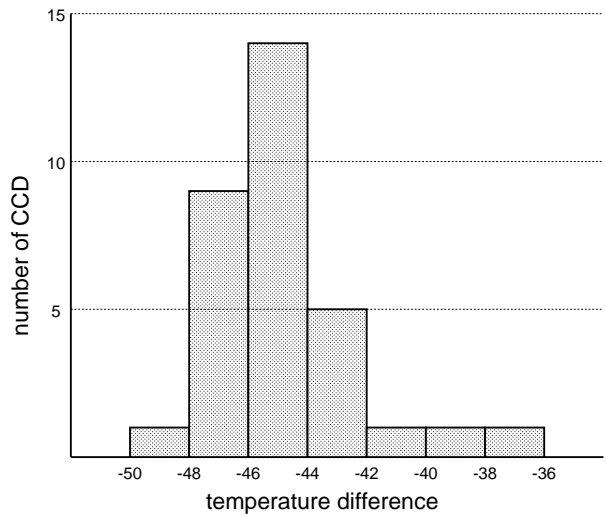}
\end{center}
\caption{Distribution of the temperature difference between CCD wafer and SSCU body.
The temperature of SSCU body is $-$21\degc, and the peltier current is 1.0~A.}
\label{fig:tec_deltaT_dist}
\end{figure}

The performance of peltier devices was investigated in the system pre-flight test,
where the whole system of MAXI including the SSCUs, the SSCE, and the DP were assembled as flight configuration.
MAXI structure was installed in a vacuum chamber, and the structure was surrounded by cooled black panels.
Radiator and LHP worked well.
Figure\,\ref{fig:tec_deltaT_dist} shows the distribution of temperature difference 
between CCD wafer and aluminum stand described in Figure\,\ref{fig:sscu}.
Peltier current was set to 1.0~A, and the temperature of the aluminum stand was $-$21\degc.
The averaged temperature difference was $-$45.4\,\degc,
which was large enough for CCDs to be kept at $<-$60\degc when SSCU bodies were at $-$20\degc.

The annealing is a good method of recovering the CCD performance degraded by charged particles (\cite{janesick2001}).
Holland \etal\ (1990) reported that the high temperature above 100\degc could significantly restore the degraded performance. 
Since we also designed that the function of LHP can be halted in the orbit by command,  
SSCUs could warm up.
The SSCU, however, must to be kept at $<$+60~\degc.
The temperature difference of $+$40~\degc between CCD wafer and SSCU body is required for the annealing at 100\degc.
We, then, tried the reverse current of peltier devices,
and confirmed that the reverse current of 0.22~A made the temperature difference of about 20\,\degc .
We, then, plan to supply the reverse current of 0.44~A for the restoration,
when the degradation of the CCD performance becomes significant after the several years of operation in the orbit. 

\subsection{Imaging and Effective Area}

\begin{figure}
\begin{center}
\FigureFile(80mm,){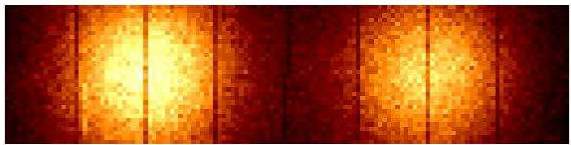}
\end{center}
\caption{An image taken with SSC-H. Mn-K X-rays from two radio-active source of $\rm ^{55}$Fe are irradiated.
Data are taken with binning=64.
The image is also binned in the horizontal direction with binning=64 by software
so that  the ratio of the horizontal axis to the vertical is the same to a real CCD configuration.}
\label{fig:ccd_img} 
\end{figure}

The X-ray direction of the SSC events is determined 
with the combination of the CCD and the CSU.
The imaging capability of the CCDs was simply checked with the $\rm^{55}$Fe source in the energy calibration.
Figure\,\ref{fig:ccd_img} is the image showing two bright spots generated by two radio active sources of $^{55}$Fe. 
We find that there is no defect region for the X-ray detection.

SSC has a large X-ray detection area,
but the effective area for the X-ray collection is limited by the CSU.
We, then, measured the effective area by using an X-ray beam line.
In the beam line, the CSU was installed in SSCU of engineering model,
and the parallel X-ray beam was irradiated into the CCD through the CSU.
As the results, we confirmed the silt width to be 2.70\,$\pm 0.05$~mm, 
which was consistent with the designed value of 2.7\,mm.
The collimator sheets reduced the effective area to 89.5\% and 86.8\% of the designed value for the SSC-H and the SSC-Z, respectively. 
It was due to the bending and non-uniform alignment of the collimator sheets.
These factors are taken into account for the astronomical analysis.

\section{Summary}

We have prepared the SSC consisting of two SSCUs.
Each sensor contains the slit, collimators, and X-ray CCDs.
The slit and collimators of each SSCU define the size of field of view to be 1$^{\circ}$.5(FWHM)$\times$90\arcdeg .
SSC has 32 CCDs of FFTCCD-4673 fabricated by Hamamatsu Photonics K.K.
The SSC has total X-ray detection area of about 200\,cm$^2$
with  a sensitive X-ray energy range of 0.5$-$12keV.
SSC operation has been optimized as a slit camera onboard MAXI. CCDs are operated in parallel-sum mode.
The binning number of 64 is standard, but it can be changed by commands.
CCDs are readout one by one in each SSCU.
The charge injection is implemented to compensate the performance degradation by charged particle.
The working temperature of CCDs is $<-$60\degc in orbit by using the peltier devices, LHP, and the radiators.
The energy resolution (FWHM) of the SSC for 5.9\,keV X-rays is 145\,eV at the CCD temperature of $-$70\degc and 149\,eV at $-$60\degc.



This work is partly supported by a Grant-in-Aid for Scientific
Research by the Ministry of Education, Culture, Sports, Science and Technology (16002004 and 19047001).  
M. K. is supported by JSPS Research Fellowship for Young Scientists (22-1677).

\bibliographystyle{elsarticle-harv}
\bibliography{<your-bib-database>}








\end{document}